# The Geometric Magnitude and Albedo for the Globe of Saturn

Anthony Mallama

14012 Lancaster Lane, Bowie, MD, 20715, USA

anthony.mallama@gmail.com

Hristo Pavlov

9 Chad Place, St. Clair, NSW 2759, Australia

hristo_dpavlov@yahoo.com

2017 January 1

Abstract

Saturn's geometric magnitude and albedo were determined by Mallama et al. (2017, Icarus 282, 19-33). That analysis depended largely on photometric data obtained when the ring system interfered with observations of the globe. In this study the magnitude and albedo are re-evaluated based on spectrophotometry obtained during a ring-plane crossing (Karkoschka, 1998, Icarus, 133, 134-146). A better correction for the illumination phase angle of Saturn's globe is also derived and it is applied to the spectroscopy. The resulting values for most of the 12 Johnson-Cousins and Sloan wavelength bands agree reasonably well with those reported by Mallama et al. but some significant differences are noted. The revised albedo value for the $R_C$ band reduces the discrepancy with that inferred from the work of Dyudina et al. (2005, ApJ 618, 973-986).

1. Introduction

The magnitudes and albedos of all the planets were determined by Mallama et al. (2017) herein referred to as Paper 1. Those quantities are believed to be very accurate for most planets. However, the Saturnian values are less certain because of interference from its ring system. The rings add reflected light of their own, obscure a portion of the globe and also cast shadows on the globe. For heuristic applications such as computing the apparent magnitude of the Saturn system for its physical ephemeris the rings must be considered. However, geophysical studies of the planet itself and analyses where Saturn is a model for exo-planets should rely on magnitudes and albedos that exclude the rings. Thus, the purpose of this paper is to re-evaluate those quantities for the globe of Saturn-only and to estimate their uncertainties. We refer to the magnitude and the albedo of Saturn observed when its disk is fully illuminated by the Sun as the 'geometric' quantities.

This paper is organized as follows. Section 2 reviews the Saturnian magnitudes derived in Paper 1 and discusses their limitations. Some of these values were derived from a combination of the spectrophotometry of Karkoschka (1998) and photometry. The spectrophotometry was obtained during a ring-plane crossing, when the rings only interfered minimally and, therefore, it is a good alternative to photometry for deriving the magnitudes and the albedos for globe of Saturn by itself. However, at the time of those observations the illumination phase angle of Saturn was near its maximum, so phase function corrections must be applied. Those functions are addressed in Section 3. The method of deriving synthetic magnitudes from spectrophotometric data is described in Section 4. The adopted phase corrections are applied to the synthetic magnitudes. Then the resulting magnitudes and albedos values are listed. Section 5 compares the results from this paper to independently derived magnitudes and albedos. The conclusions of this study are summarized in Section 6. An appendix describes validation of the observational data and its processing.

2. Discussion of magnitudes

Paper 1 reported the magnitudes of Saturn for the 12 most common filters currently employed in wide-band photometry. The 7 Johnson-Cousins bands (U, B, V, R, I, $R_C$ and $I_C$) and the 5 principal Sloan bands (u', g', r', i', and z') cover the spectrum from about 0.3 to 1.0 μm. Albedo values were then derived from magnitudes of Saturn for each band.

The Johnson-Cousins magnitudes of Saturn in the U, B, V, R and I bands were taken from an earlier photometric analysis by Mallama (2012). That study solved for the integrated brightness of Saturn including its ring system based on observations from the Earth taken over a wide range of illumination phase angles and ring inclination angles. The inclination is highly important because it affects the integrated brightness more strongly than does the small range of phase angles seen from the Earth. The five heuristic formulas for the integrated magnitudes were intended primarily for physical ephemeris calculations (such as those found in almanacs) and they are believed to be the best available. However, the formulas are limited in their ability to represent the brightness of the globe of Saturn without the rings for the purposes of precise geophysical investigations of the planet.

The remaining Johnson-Cousins magnitudes ($R_C$ and $I_C$) were synthesized from the spectrophotometry of Karkoschka (1998) alone because photometry is lacking in these bands. (The method of synthesizing magnitudes is described in Section 4 of this paper.) Karkoschka's data were acquired when the Earth was only 0.6 degrees below the ring-plane and the Sun was 2.0 degrees above. His narrative discusses correction for the remaining small effect of occultation and eclipse of the globe by the ring, and he states that 'the given spectrum for Saturn refers to the case of Saturn without rings'. The phase angle at the time of these observations was 5.7 degrees and Karkoscha recommended that 5% should be added to the spectrum values. The size of this correction will be discussed in the next section.

The Sloan magnitudes were an average of newly acquired photometric results, synthetic magnitudes and values transformed from the Johnson-Cousins system. The Sloan photometry was obtained over a limited period of time that did not allow for sampling and analysis of the effects of phase angle and inclination variations. Therefore, the Johnson-Cousins phase and inclination coefficients were mapped to Sloan coefficients by interpolation in wavelength and those values were applied to the photometry in order to determine the geometric magnitude of Saturn's disk without the rings. Synthetic magnitudes were derived from Karkoschka's spectroscopy and the same issue noted above regarding correction for the phase angle applies. The method of transformation from Johnson-Cousins magnitudes to the Sloan system is described in Paper 1 and depends on the equations listed by Taylor (1986), Fukugita et al. (1996) and Smith et al. (2002).

Thus, the geometric albedos were derived from corresponding magnitudes which, in turn, were determined in several different ways according to the availability and quality of photometric and spectroscopic data. In this paper only ring-plane crossing spectrophotometry will be employed. The required phase angle corrections to that data are addressed in the next section.

3. Phase function

A model for the combined magnitude of Saturn and its ring was derived by Mallama (2012) and subsequently reported in Paper 1 by fitting photometric results to Equation 1

$$M_1(\alpha,\beta) = C_0 + C_1 \sin(\beta) + C_2 \alpha - C_3 \sin(\beta) e^{(C_4 \alpha)}$$

Equation 1

where $M_1$ is the magnitude at one astronomical unit, $\alpha$ is the illumination phase angle, $\beta$ is the inclination angle of the ring system. Thus, $M_1(0,0)$ is the geometric magnitude. The Cs are the fitted coefficients.

The derived coefficients indicate that the effect of the inclination and phase angle of the rings dominates that of phase angle of the globe. Mallama (2012) reports that "When Saturn is opposite the Sun and its rings are at their maximum inclination the system is 0.99 magnitude brighter in V than when the rings are edge-on. The equivalent luminosity ratio, 2.49, indicates that the rings contribute nearly 150% as much brightness as the planet. When at quadrature the magnitude difference is 0.82 and the ring's contribution is only about 110% as great as the planet." Thus, the inclination of the rings is a critical factor in modeling the system of Saturn and its rings. Moreover, the phase angle affects the brightness of the rings much more strongly than it affects the globe as shown in Figure 1.

The phase function of the globe alone, according to this model, may be evaluated from Equation 1 by setting β to zero. If $C_0$ is also set to zero (for normalization) then magnitude is simply the product of $C_2$ and α. Table 1 lists the $C_2$ coefficients and the amount of dimming of Saturn at the 5.7 degree phase angle the spectrophotometry(Karkoschka, 1998) relative to that at phase angle zero. These values are quite large and are unrealistic for the globe by itself.

The last line of Table 1 shows the corresponding dimming for Jupiter at the same phase angle based on the phase function coefficients for that planet reported by Mallama and Schmude (2012). The phase curve for the globe of Jupiter is believed to be very accurate because there is no ring system to interfere. The table indicates much less dimming in the case of Jupiter. Since the phase curve for Jupiter spans angles from 0 to 12 degrees it may be applied as a proxy for the Saturnian phase function at least to determine the geometric magnitude and albedo from observations made at small phase angles.

Another function that deserves consideration is that derived by Dyudina et al. (2005) which was determined from earlier photometric models of Saturn based, in turn, on Pioneer spacecraft imagery. Those earlier models were the anisotropic scattering functions derived by Tomasko and Doose (1984) and Dones et al. (1993) which were originally applied to characterization of the planet's atmosphere. Dyudina et al. reused the parameters of the fits to derive integrated planetary brightness as a function of the orbital phase angle for an exo-planet similar to Saturn. Orbital phase angles at the vast distance of exo-planets are essentially the same as the illumination phase angles for solar system bodies and in the case of angle zero they are identical.

Table 1. Phase angle coefficients and dimming

|  | U | B | V | R | I |
|---|---|---|---|---|---|
| Coefficient $C_2$ (Saturn)[1] | 0.020 | 0.034 | 0.026 | 0.032 | 0.032 |
| Dimming at 5.7 degrees (Saturn) | 11.1% | 19.5% | 14.6% | 18.3% | 18.3% |
| Dimming at 5.7 degrees (Jupiter) | 4.5% | 1.5% | 1.6% | 1.0% | 1.1% |

[1] magnitudes per degree of phase angle

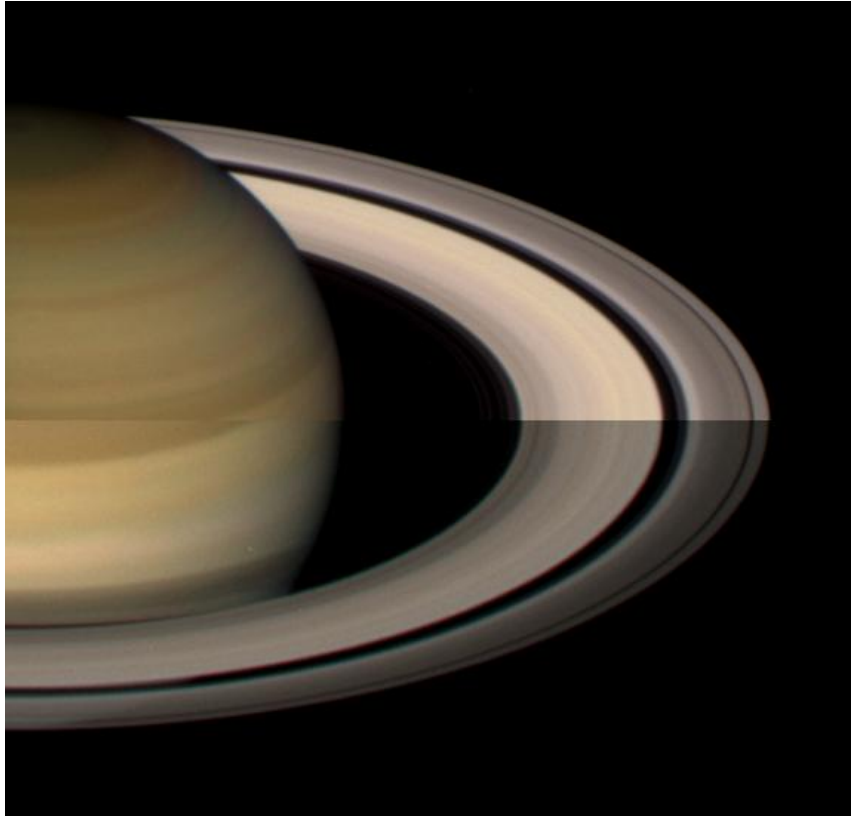

Figure 1. This composite image of Saturn and its rings shows the enhanced brightness of the rings near phase angle 0 degrees (top half) as compared to phase angle 6 degrees (bottom half). The color composite was generated from Hubble Space Telescope images taken through the F439W (blue), F555W (green) and F675W (red) filters. The top half is from images U97F1105m, U97F1109m and U97F110em and the bottom from U639A102r, U639A103r and U639A104r. The data were retrieved from the MAST archive. This figure was originally published by Mallama (2012).

Figure 7 of Dyudina et al. illustrates the phase function for the realistic case of a 10% oblate Saturn as seen from the equator in the red Pioneer filter spanning 0.6 to 0.7 μm. There is no numerical data to accompany this plot, so their phase function was determined by digitizing the figure. The dimming at phase angle 5.7 degrees was found to be about 1.5%. This very small value suggest that the minimal dimming at wavelengths longer than the UV as indicated by the Jupiter model (Table 1) also applies reasonably well to Saturn.

The adopted phase angle correction in this paper is 1.5% for visible and near-IR bands as indicated by the Jupiter model in Table 1 and by Dyudina et al. Since all those values are quite consistent the uncertainty is generously taken to be 1%. The adopted correction for the U and u' bands is 5% as indicated by Table 1 and by Karkoschka. This value is not very well determined and so its uncertainty is taken as 3%.

4. Magnitudes and albedos from spectroscopy only

As discussed in Section 2, the spectrophotometry of Karkoschka (1998) was obtained when Saturn's ring system only interfered minimally. On the other hand, most of the photometric data was taken when the ring system contributed a large portion of the total light being measured. Therefore we adopt the spectroscopic data as the basis for the magnitudes and albedos in this study.

Synthetic magnitudes were derived from the spectroscopy by integrating the product of spectral flux and instrumental response over the frequency range of each band. The relationship is indicated by Equation 2 (Smith et al., 2002 and Fukugita et al., 1996), where $v$ is frequency, $S_v$ is the system response and $f_v$ is flux at the top of the atmosphere in ergs/s/cm$^2$/Hz.

$$m = -2.5 \frac{\int d(\log v)\ f_v\ S_v}{\int d(\log v)\ S_v}$$

Equation 2

The resulting synthetic magnitudes are listed in the rows labeled 'raw magnitude' in the Johnson-Cousins and the Sloan portions of Table 2. The 'geometric magnitude' in the following rows accounts for the phase angle corrections adopted in the previous section of this paper. The 'magnitude uncertainty' takes into account a 4% uncertainty given by Karkoschka for the absolute flux of his spectrophotometry as well as the 1.5% and 3% phase function uncertainties given in the previous section. These rows are followed by the corresponding geometric albedo and its uncertainty. Then the geometric albedo values from Paper 1 are listed. These are followed by the differences between the two albedos in absolute units and in uncertainty ('sigma') units. Finally, the effective wavelength and full width at half maximum for each band is listed.

The average of the albedo differences taken across all 12 bands, in the sense 'this paper' minus 'Paper 1', is -0.014 and the corresponding root-mean-square (RMS) difference is 0.042. Thus, the original and revised values agree reasonably well in general. Figure 2 illustrates the albedos and their differences.

Table 2. Synthetic magnitudes, geometric albedos and band-pass characteristics

```
Johnson-Cousins system
```

|  | U | B | V | R | I | Rc | Ic |
|---|---|---|---|---|---|---|---|
| Raw magnitude | -7.20 | -7.82 | -8.93 | -9.57 | -9.65 | -9.40 | -9.55 |
| Geometric magnitude | -7.25 | -7.84 | -8.95 | -9.59 | -9.66 | -9.42 | -9.57 |
| Magnitude uncertainty | 0.05 | 0.04 | 0.04 | 0.04 | 0.04 | 0.04 | 0.04 |
| Geometric albedo | 0.238 | 0.339 | 0.517 | 0.568 | 0.444 | 0.553 | 0.464 |
| Albedo uncertainty | 0.012 | 0.014 | 0.021 | 0.023 | 0.018 | 0.023 | 0.019 |
| Albedo (Paper 1) | 0.203 | 0.339 | 0.499 | 0.568 | 0.423 | 0.646 | 0.543 |
| Difference (absolute) | 0.035 | 0.000 | 0.018 | 0.000 | 0.021 | -0.093 | -0.079 |
| Difference (sigmas) | 3.0 | 0.0 | 0.8 | 0.0 | 1.1 | -4.1 | -4.1 |
| Effective wavelength[1] | 0.360 | 0.436 | 0.549 | 0.700 | 0.900 | 0.641 | 0.798 |
| Full-width half max.[1] | 0.068 | 0.098 | 0.086 | 0.209 | 0.221 | 0.158 | 0.154 |

```
Sloan system
```

|  | u' | g' | r' | i' | z' |
|---|---|---|---|---|---|
| Raw magnitude | -6.34 | -8.39 | -9.22 | -9.24 | -8.88 |
| Geometric magnitude | -6.39 | -8.41 | -9.23 | -9.26 | -8.90 |
| Magnitude uncertainty | 0.05 | 0.04 | 0.04 | 0.04 | 0.04 |
| Geometric albedo | 0.242 | 0.413 | 0.590 | 0.547 | 0.403 |
| Albedo uncertainty | 0.012 | 0.017 | 0.024 | 0.023 | 0.017 |
| Albedo (Paper 1) | 0.209 | 0.436 | 0.618 | 0.553[2] | 0.450 |
| Difference (absolute) | 0.033 | -0.023 | -0.028 | -0.006 | -0.047 |
| Difference (sigmas) | 2.7 | -1.3 | -1.2 | -0.3 | -2.8 |
| Effective wavelength[1] | 0.355 | 0.469 | 0.616 | 0.748 | 0.893 |
| Full-width half max.[1] | 0.063 | 0.143 | 0.140 | 0.149 | 0.117 |

[1] units of microns

[2] this value was incorrectly listed as 0.601 in Paper 1

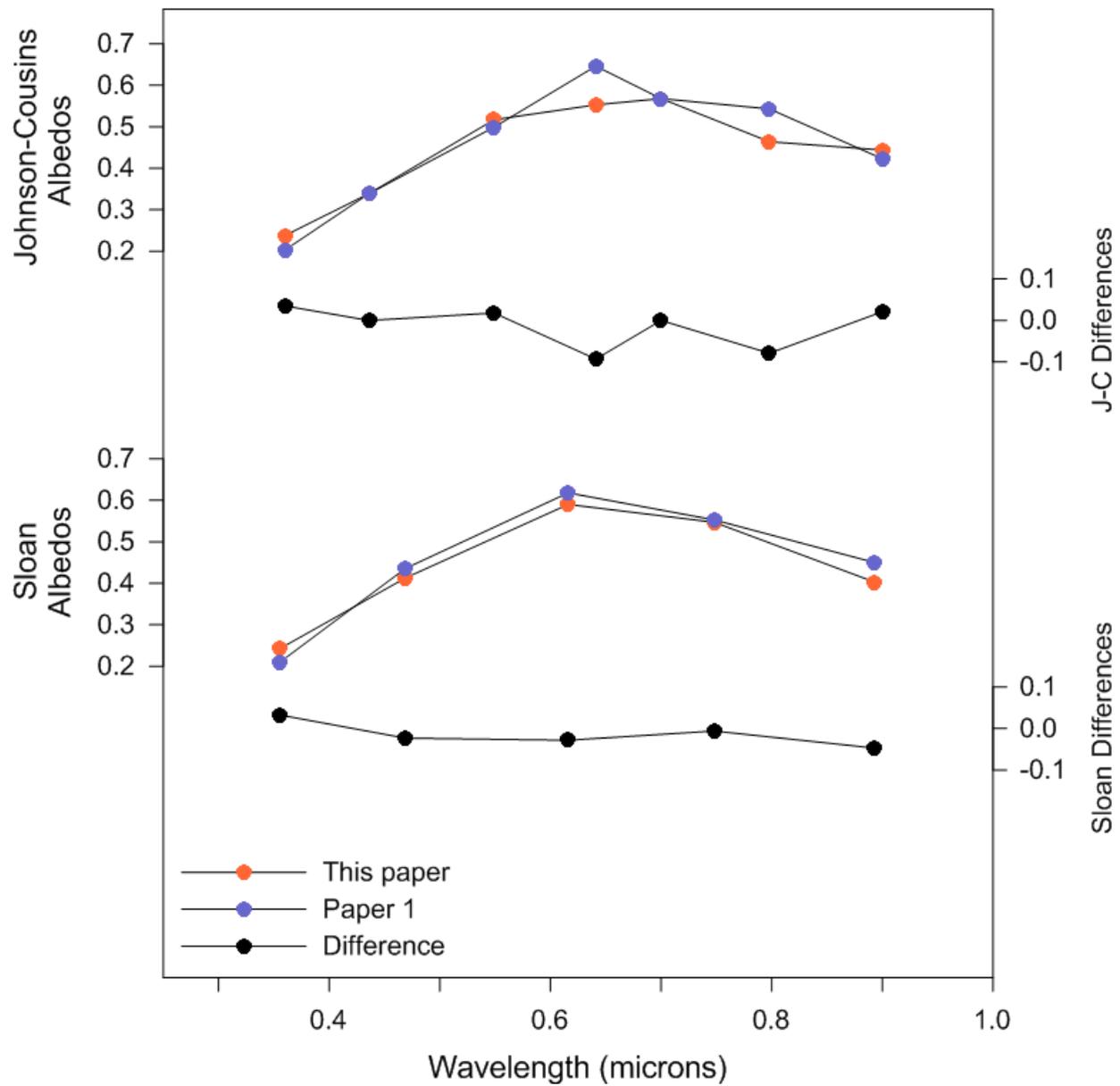

Figure 2. Albedo values from this paper and from Paper 1 are compared.

5. Comparisons with other magnitude and albedo determinations

A widely accepted source of astrophysical reference data is that of Cox (2000). The geometric V magnitude of Saturn given there in Table 12.8 is -8.88. By comparison the magnitude from Table 2 in this paper is 7% brighter and that given Paper 1 is 3% brighter. Cox lists the B-V value as 1.04, resulting in a B magnitude of -7.84 which is exactly equal to the values in Table 2 and Paper 1. The column listing the color index involving the U band is labeled by Cox as 'U – V' but that is surely meant to be 'U – B' and its value is 0.58. So Cox's U magnitude would be -7.26 which is 1% brighter than the value in Table 2 and 18% brighter than Paper 1. (If 0.58 was actually the U – V color index then Cox's U magnitude would be -8.30 which is 161% brighter than the value in Table 2 and 208% brighter than Paper 1.)

Cox also lists the 'visual geometric albedo' of Saturn which is given as 0.47. The value in Table 2 is 10% brighter and that from Paper 1 is 6% brighter. These percentage differences in albedos are greater than the corresponding magnitude differences by 4% and 3%, respectively. One possible reason is that different solar magnitudes were used and another is that Cox's albedo value corresponds to something other than the V band.

Despite the different methods and data sources employed in this study and in Paper 1, the magnitudes and albedos in most of the 12 wavelength bands did not change significantly. The Johnson-Cousins B, V, R and I results differed by only about one sigma and likewise with those for the Sloan g', r' and i' bands. The larger differences for the two ultraviolet bands, U and u', are not too surprising given that the photometric and spectroscopic observations were made from the ground where atmospheric extinction is very high. The $I_C$ and z' differences in the IR cannot be explained in the same way though. One possible explanation is that Saturn's IR brightness has changed. It is noteworthy that the IR brightness of Neptune increased substantially over time as described by Schmude et al. (2016).

The smaller albedo value derived for the $R_C$ band in this paper is the most interesting change because it may help to resolve a discrepancy. Paper 1 compared the $R_C$ band (0.56 – 0.72 µm) albedo to that derived from Figure 6 of Dyudina et al. (0.6 – 0.7 µm). The $R_C$ value was found to be 50% larger. The authors of Paper 1 speculated that neither their value nor that of Dyudina et al. was very accurate and that the correct albedo is probably between the two extremes. (See Section 3 of this paper for more information about the analysis of Figure 6 of Dyudina et al.)

The revised $R_C$ albedo of 0.553 in Table 2 reduces the difference with Dyudina et al. to 28%. The remaining difference may be due to the averaging of the scattering properties of Saturnian zones and belts (the two types of atmospheric bands) by Dyudina et al. As discussed in Paper 1 simple averaging may result in too low an overall albedo value because the bright zones span a much larger latitude range than do the darker belts. Figure 3 of Mallama (2014) shows that the extent of the zones far exceeds that of the belts. Another possibility is that the albedo of Saturn was actually lower when Pioneer imaged the planet in 1979.

6. Conclusions

The presence of Saturn's ring system complicates the determination of the magnitudes and albedos of the planetary globe itself. Therefore, the values reported in Paper 1 are re-derived based solely on spectrophotometry obtained during the most recent ring-plane crossing (Karkoschka 1998) when the rings only interfered minimally.

The illumination phase function is also reexamined. The phase function of Jupiter (Mallama and Schmude, 2012), that function derived for Saturn from the work of Dyudina et al. (2005) and the magnitude corrections recommended by Karkoscha are considered. As a result, smaller corrections are applied to Karkoschka's spectrophotometry than were made in Paper 1.

Even with these changes, most of the magnitudes and albedos did not change significantly. The larger differences found for the U, u', $R_C$, $I_C$ and z' bands were discussed. The change to $R_C$ brings the albedo derived in this paper closer to that inferred from the work of Dyudina et al.

Appendix – Further validation

The results reported in this paper depend upon synthetic magnitudes derived from the spectroscopy of Karkoschka (1998). So, this appendix addresses the accuracy of the flux itself and that of the synthetic magnitudes which were derived from that flux.

Karkoschka states that the absolute accuracy for his spectrophotometry is about 4%. This estimated uncertainty was verified by comparison with Hubble Space Telescope observations. A search of the MAST archive of HST Wide Field Planetary Camera 2 data for images taken during the ring-plane crossings of 1995 located observations recorded on August 6 and November 17 and 18. The November data was chosen for comparison with Karkoschka's spectrophotometry because it was acquired at nearly the same phase angle as the spectrophotometry (5.4 and 5.7 degrees, respectively). The August data taken at 3.9 degrees was not used.

Photometry was derived from the HST data using the Aperture Photometry Tool (Laher et al. 2012). An elliptical aperture was fitted to the globe of Saturn with a surrounding annulus for background subtraction. The energy within the aperture was taken to be the product of the recorded counts and the inverse sensitivity constant *PHOTFLAM* in the image header. The corresponding energy for the Sun was taken from Wehrli's table (http://rredc.nrel.gov/solar/spectra/am0/) at the wavelength indicated by *PHOTPLAM* in the image header. Then the distances from Saturn to the Earth and to the Sun as well as the area subtended by Saturn's disk were taken into account in determining albedo values. The area of the ring's shadow on the disk of the planet was measured to be 2.5% and that amount was added to the albedo.

Figure 3 illustrates good agreement between the spectroscopy and the photometry. A quantitative comparison beyond 600 nm is difficult because the HST filters partially coincide with steep and narrow absorption bands. Likewise the 336 nm filter is problematic due to the uncertainty of the ground-based spectrophotometry taken at UV wavelengths where atmospheric extinction is high. A comparison of the data points between 400 and 600 nm indicates a mean albedo difference of +0.009 in the sense 'Karkoschka minus HST' with an RMS difference of 0.018. When these same statistics are computed from percentage differences the mean is +2.3% and the RMS is 4.1%. Thus, Karkoschka's estimated uncertainty of 4% is validated at least in the visible part of the spectrum.

The synthetic magnitudes in this paper were checked by comparison with the U, B and V values derived by Karkoschka from the same data. While the results should be exactly the same in principle there are reasons why they might be slightly different. Two examples are the method of computation and the response curves for the filters. Table 3 shows that there are, in fact, differences between the magnitudes from Table 2 of this paper and those listed by Karkoschka. The values in Table 2 average 0.06 brighter with the standard deviation of the mean equal to 0.04.

The procedure for computing synthetic magnitudes used in this paper, described in Section 4, was previously employed by Mallama (2015) to validate the correspondence between HST stellar CalSpec spectrophotometry (Bohlin et al. 2014) and Sloan system standard star magnitudes (Smith et al. 2002) at a level of about 0.01 magnitude. So it is known to be a reliable method. Nevertheless the difference with Karkoschka prompted testing with a second method for computing synthetic magnitudes. In this approach the planetary albedo is converted to absolute flux based on the Hubble Space Telescope CalSpec flux of the Sun and on the area of Saturn's disk. Then the computed absolute flux of Saturn and the CalSpec absolute flux of Vega are integrated over the corresponding filter response curve to derive a total flux for each filter. Finally the synthetic magnitude is computed as the magnitude relative to that of Vega.

The results of the second method are shown in the last line of Table 3 where the mean difference is found to be 0.05 brighter than Karkoschka (which is very similar to 'Table 2 minus Karkoschka') and the standard deviation of the mean difference is 0.07. The mean difference between the second method and the values in Table 2 of this paper for U, B and V is +0.01 magnitude with a standard deviation of the mean equal to 0.05. So, the methods used in this paper give results that are mutually consistent on the average even if they are slightly brighter than those of Karkoscka.

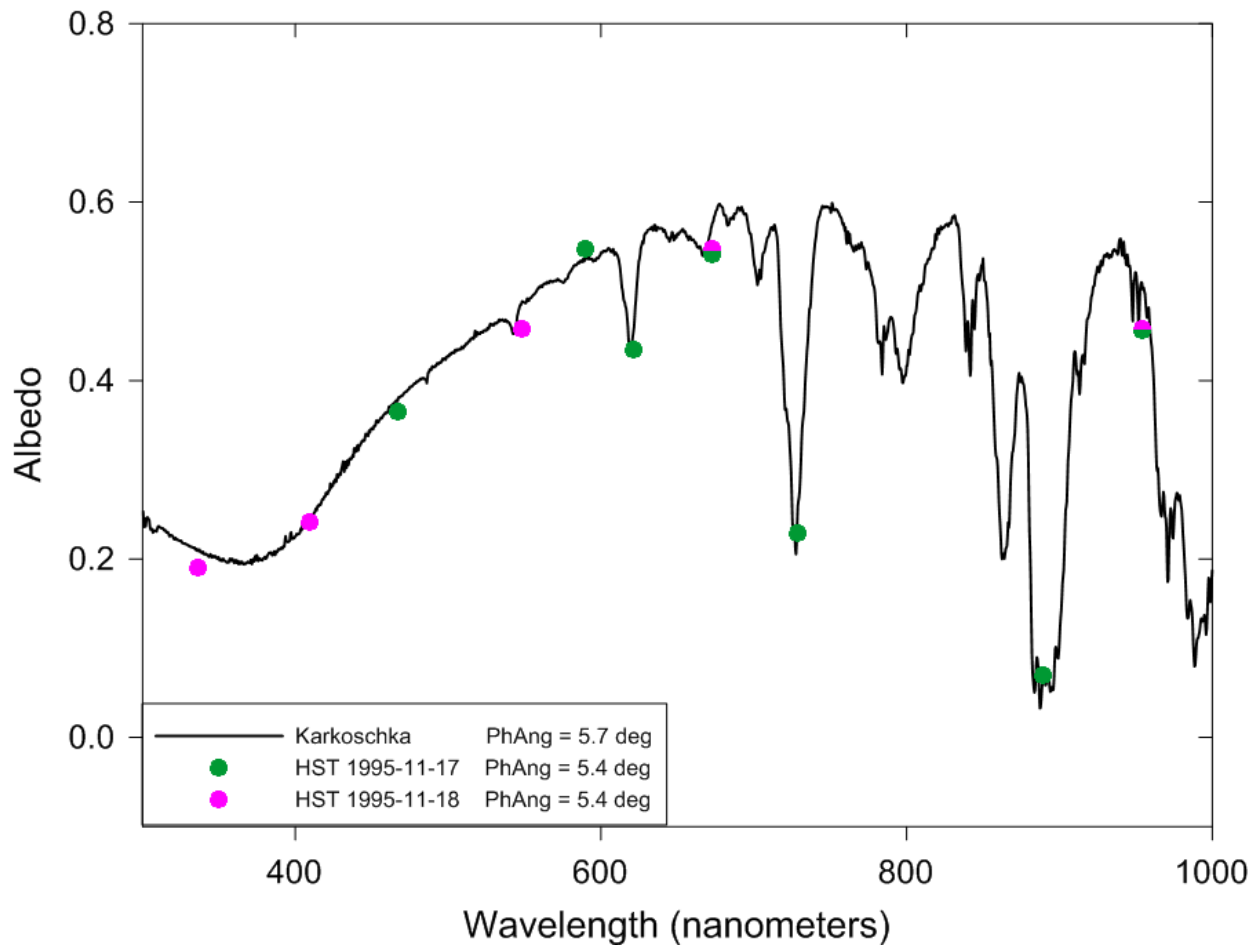

Figure 3. The comparison between albedo values derived on HST photometry and those reported by Karkoschka indicates good agreement.

Table 3. Synthetic magnitudes compared

```
                                                      Mean      St.D.
                              U       B       V       Diff.    of Mean
Karkoschka (1998)           -7.12   -7.80   -8.87      ---        ---
This paper (from Table 2)   -7.20   -7.82   -8.93     -0.06      0.04
This paper (second method)  -7.24   -7.84   -8.86     -0.05      0.05
```